\documentclass[11pt,leqno]{trans2-l}
\usepackage{amsmath,amsthm,amssymb,amsxtra}
\usepackage[all]{xy}

\let\geq\geqslant
\let\leq\leqslant

\makeatletter
\renewcommand\fnum@table{\tablename\ }

\long\def\@makecaption#1#2{%
  \setbox\@tempboxa\vbox{\color@setgroup
    \advance\hsize-2\captionindent\noindent
    \@captionfont\@captionheadfont#1\@xp\@ifnotempty\@xp
        {\@cdr#2\@nil}{\@captionfont\upshape #2.}%
    \unskip\kern-2\captionindent\par
    \global\setbox\@ne\lastbox\color@endgroup}%
  \ifhbox\@ne 
    \setbox\@ne\hbox{\unhbox\@ne\unskip\unskip\unpenalty\unkern}%
  \fi
  \ifdim\wd\@tempboxa=\z@ 
    \setbox\@ne\hbox to\columnwidth{\hss\kern-2\captionindent\box\@ne\hss}%
  \else 
    \setbox\@ne\vbox{\unvbox\@tempboxa\parskip\z@skip
        \noindent\unhbox\@ne\advance\hsize-2\captionindent\par}%
  \fi
  \ifnum\@tempcnta<64 
    \addvspace\abovecaptionskip
    \hbox to\hsize{\kern\captionindent\box\@ne\hss}%
  \else 
    \hbox to\hsize{\kern\captionindent\box\@ne\hss}%
    \nobreak
    \vskip\belowcaptionskip
  \fi
\relax
}

\makeatother

\newcommand{\st}[1]{\ensuremath{^{\scriptstyle \textrm{#1}}}}

\newcommand\bigcheck[1]{#1 \raise1ex\hbox{$\hspace{-1ex}{}^\vee$}}
\newcommand\sucheck[1]{#1 \raise0.5ex\hbox{$\hspace{-1ex}{}^\vee$}}

\newcommand{\romanparenlist}{
  \renewcommand{\theenumi}{\roman{enumi}}%
  \renewcommand{\labelenumi}{(\theenumi)}%
}

\newcommand{\ad}{\mathop{\rm ad}\,}

\renewcommand{\ne}{\mathop{\rm ne}\,}

\newcommand{\Pyr}{{\rm Pyr}}
\newcommand{\rank}{{\rm rank}}

\renewcommand{\sl}{s\ell}

\newcommand{\FF}{\mathbb{F}}
\newcommand{\NN}{\mathbb{N}}

\newcommand{\RR}{\mathbb{R}}
\newcommand{\ZZ}{\mathbb{Z}}

\newcommand{\fg}{\mathfrak{g}}
\newcommand{\fh}{\mathfrak{h}}
\newcommand{\fl}{\mathfrak{l}}

\newcommand{\fo}{\mathfrak{o}}
\newcommand{\fp}{\mathfrak{p}}
\newcommand{\fs}{\mathfrak{s}}
\newcommand{\ft}{\mathfrak{t}}

\newcommand{\fgl}{\mathfrak{gl}}
\newcommand{\fsl}{\mathfrak{sl}}
\newcommand{\fso}{{\mathfrak{s}}\mathfrak{o}}
\newcommand{\fsp}{{\mathfrak{s}}\mathfrak{p}}

\makeatletter
\renewcommand\section{\@startsection {section}{1}{\z@}%
                                   {-3.5ex \@plus -1ex \@minus -.2ex}%
                                   {2.3ex \@plus.2ex}%
                                   {\normalfont\large\bfseries}}
\renewcommand\subsection{\@startsection{subsection}{2}{\z@}%
                                     {-3.25ex\@plus -1ex \@minus -.2ex}%
                                     {0ex \@plus .0ex}%
                                     {\normalfont\normalsize\bfseries}}

\@addtoreset{equation}{section}
\makeatother

\newtheorem{theorem}{Theorem}[section]
\newtheorem{lemma}{Lemma}[section]
\newtheorem{corollary}{Corollary}[section]
\newtheorem{proposition}{Proposition}[section]
\newtheorem*{lemma*}{Lemma}

\theoremstyle{definition}

\theoremstyle{remark}
\newtheorem{remark}{Remark}[section]

\begin{document}

\title[Good Gradings of Simple lie algebras]{Classification of
  Good Gradings of Simple Lie Algebras}

\author{A.G. Elashvili}
\address{Razmadze Matematics Institute\\
  1 M. Aleksidze str. 0193 Tbilisi\\
  Republic of Georgia}
\email{alela@rmi.acnet.ge}
\thanks{Supported in part by SFB/TR 12 at
    the Ruhr-University Bochum}

\author{V.G. Kac}
\address{Department of Mathematics\\
  M.I.T.\\
  Cambridge, MA 02139, USA}
\email{kac@math.mit.edu}
\thanks{Supported in
    part by NSF grant DMS-0201017.}

\begin{abstract}
We study and give a complete classification of good
$\ZZ$-gradings of all simple finite-dimensional Lie algebras.
This problem arose in the quantum Hamiltonian reduction for
affine Lie algebras.

\end{abstract}

\maketitle

\section{Introduction}
\label{sec:intro}

Let $\fg$ be a finite-dimensional Lie algebra over an
algebraically closed field $\FF$ of characteristic~$0$.  Let $\fg =
\oplus_{j \in \ZZ} \fg_j$ be a $\ZZ$-grading of $\fg$
(i.e.,~$[\fg_i, \fg_j] \subset \fg_{i+j}$), and let $\fg_+
=\oplus_{j>0} \fg_j$, $\fg_{\geq}= \oplus_{j\geq0} \fg_j$.

An element $e \in \fg_2$ is called \emph{good} if the following
properties hold:
\begin{eqnarray}
\label{eq:0.1}
  \ad e: \fg_j \to \fg_{j+2} \hbox{ is injective for } j\leq -1\, ,\\
\label{eq:0.2}
  \ad e : \fg_j \to \fg_{j+2} \hbox{ is surjective for } j \geq -1 \, .
\end{eqnarray}
Obviously, $e$ is a non-zero nilpotent element of $\fg$.  Note
that (\ref{eq:0.1}) is equivalent to
\begin{equation}
  \label{eq:0.3}
   \hbox{the centralizer } \fg^e \hbox{ of }e \hbox{ lies in } \fg_{\geq} \, .
\end{equation}
 Also, (\ref{eq:0.1}) and
(\ref{eq:0.2}) for $j=-1$ imply that
\begin{equation}
  \label{eq:0.4}
  \ad e : \fg_{-1} \to \fg_1 \hbox{ is bijective.}
\end{equation}
Finally, (\ref{eq:0.2}) for $j=0$ means that
\begin{equation}
  \label{eq:0.5}
  [\fg_0 ,e]=\fg_2 \, .
\end{equation}

Denote by $G$ the adjoint group corresponding to the Lie algebra
$\fg$ and by $G_0$ its subgroup consisting of the elements
preserving the $\ZZ$-grading.  Then (\ref{eq:0.5}) implies that
$G_0 \cdot e$  is a Zariski dense open orbit in $\fg_2$.
Consequently, all good elements form a single $G_0$-orbit in
$\fg_2$, which is Zariski open.

A  $\ZZ$-grading of $\fg$ is called \emph{good} if it admits a
good element.

The most important examples of good $\ZZ$-gradings of $\fg$
correspond to $\sl_2$-triples $\{ e,h,f \}$, where $[e,f]=h$,
$[h,e]=2e$, $[h,f]=-2f$.  It follows from representation theory of
$\sl_2$ that the eigenspace decomposition of $\ad \, h$ in $\fg$
is a $\ZZ$-grading of $\fg$ with a good element $e$.  We call the
good $\ZZ$-gradings thus obtained the \emph{Dynkin
  $\ZZ$-gradings}.

In the present paper we classify all good $\ZZ$-gradings of simple
Lie algebras.  More precisely, for each nilpotent element $e$ of
a simple Lie algebra $\fg$, we find all good $
\ZZ$-gradings of $\fg$ for which $e$ is a good element.  This
problem arose in the study of a family of
vertex algebras obtained from an affine (super)algebra, associated
to a simple finite-dimensional Lie (super)algebra $\fg$ with a
non-degenerate invariant bilinear form, by the quantum Hamiltonian
reduction (cf.~\cite{FORTW}, \cite{KRW}, \cite{KW}). The method
developed in the present paper works in the ``super'' case as
well. We would like also to point out that the notion of a good
grading helps to prove new results even for the well studied case
of Dynkin gradings, like Theorem~\ref{th:1.5}.

The first named author is grateful to V.~Ginzburg for the
reference \cite{FORTW}, to D.~Panyushev for useful discussions,
and especially to M.~Jibladze, the author of
Proposition~\ref{prop:4.1}, for being a patient listener. We are
grateful to K. Baur and N. Wallach for pointing out in
math.RT/0409295 an omission in Theorem~\ref{th:6.4}.

\section{Properties of good gradings}
\label{sec:1}

{}From now on, we shall assume that $\fg$ is a semisimple Lie
algebra.  Fix a $\ZZ$-grading of $
\fg$:
\begin{equation}
  \label{eq:1.1}
  \fg = \oplus_{j \in \ZZ} \fg_j\, .
\end{equation}

\begin{lemma}
  \label{lem:1.1}
Let  $e \in \fg_2$, $e \neq 0$.  Then there exists $h \in \fg_0$ and $f
\in \fg_{-1}$ such that $\{ e,h,f \}$ form an $\sl_2$-triple,
i.e., $[h,e]=2e$, $[e,f]=h$, $[h,f]=-2f$.
\end{lemma}

\begin{proof}
  By the Jacobson--Morozov theorem (see  \cite{J}), there
  exist $h,f \in \fg$ such that $\{ e,h,f\}$ is an
  $\sl_2$-triple.  We write $h=\sum_{j \in \ZZ} h_j$, $f=\sum_{j
    \in \ZZ}f_j$ according to the given $\ZZ$-grading of
  $\fg$.  Then $[h_0,e]=2e$ and $[e,\fg] \ni h_0$ (since
  $[e,f_{-2}]=h_0$).  Therefore, by Morozov's lemma (see
   \cite{J}), there exists $f'$ such that $\{ e,h,f'\}$ is an
  $\sl_2$-triple.  But then $\{ e,h_0,f'_0 \}$ is an $\sl_2$-triple.
\end{proof}

The following lemma is well-known \cite{C} (and easy to prove).

\begin{lemma}
  \label{lem:1.2}
Let $e$ be a non-zero nilpotent element of $\fg$ and let $\fs=\{
e,h,f \}$ be an $\sl_2$-triple.  Then $\fg^{\fs}$ (the
centralizer of $\fs$ in $\fg$ ) is a maximal reductive
subalgebra of $\fg^e$.
\end{lemma}

\begin{theorem}
  \label{th:1.1}
Let (\ref{eq:1.1}) be a good $\ZZ$-grading and
$e \in \fg_2$ a good element.  Let $H \in \fg$ be the element
defining the $\ZZ$-grading (i.e., $\fg_j = \{ a \in \fg | [H,a]=ja
\}$), and let $\fs =\{ e,h,f \}$ be an $\sl_2$-triple given by
Lemma~\ref{lem:1.1}.  Then $z:=H-h$ lies in the center of
$\fg^{\fs}$.
\end{theorem}

\begin{proof}
The element $H$ exists since all derivations of $\fg$ are inner.   By (\ref{eq:0.3}) the eigenvalues of $\ad H$ on $\fg^e$ are
  non-negative.  Hence the eigenvalues of $\ad H$ on $\fg^{\fs}$ are
  non-negative.  Since by Lemma~\ref{lem:1.2}, $\fg^{\fs}$ is
  reductive,  it follows that $[H,\fg^{\fs}] =0$.  But, obviously,
  $[h,\fg^{\fs}]=0$, hence $[z,\fg^{\fs}]=0$.
\end{proof}

\begin{corollary}
  \label{cor:1.1}
If $\fs=\{e,h,f\}$ is an $\sl_2$-triple in $\fg$ and the center of
$\fg^{\fs}$ is trivial, then the only good grading for which $e$
is a good element is the Dynkin grading.
\end{corollary}

It is well known that $\fg_0$ is a reductive subalgebra of $\fg$ and
a Cartan subalgebra $\fh$ of $\fg_0$ is a Cartan subalgebra of
$\fg$. Let $\fg=\fh\oplus(\bigoplus_\alpha\fg_\alpha)$ be the
root space
decomposition of $\fg$ with respect to $\fh$. Let  $\Delta_0^+$ be
a system of positive roots of the subalgebra $\fg_0$. It is well
known that $\Delta^+=\Delta_0^+\cup\left(\alpha\ \,| \,\ \fg_\alpha\subset\fg_s,\
s>0\right)$ is a set of positive roots of $\fg$. Let
$\Pi = \{ \alpha_1 , \ldots ,\alpha_r \}\subset \Delta^+$ be the set of simple roots. Setting
$\Pi_s=(\alpha\in\Pi|\fg_\alpha\subset\fg_s)$ we obtain a
decomposition of $\Pi$ into a disjoint union of subsets
$\Pi=\cup_{s\geq0}\Pi_s$. This decomposition is called the
\emph{characteristic} of the $\ZZ$-grading (\ref{eq:1.1}). So we obtain a bijection
between all $\ZZ$-gradings of $\fg$ up to conjugation and all
characteristics.  A $\ZZ$-grading is called \emph{even} if its
characteristic is $\Pi = \Pi_0 \cup \Pi_2$.

\begin{theorem}
  \label{th:1.2}
If the $\ZZ$-grading (\ref{eq:1.1})  is good, then $\Pi = \Pi_0
\cup \Pi_1 \cup \Pi_2$.
\end{theorem}

\begin{proof}
  Let $e \in \fg_2$ be a good element, and suppose that $\alpha_j
  \notin \Pi_0 \cup \Pi_1 \cup \Pi_2$
  for some $j$.  Then $e$ lies in the subalgebra generated by the
  $e_{\alpha_i}$, $i \neq j$, and therefore
  $[e,e_{-\alpha_j}]=0$.  This contradicts (\ref{eq:0.1}).
\end{proof}

This  result was proved by Dynkin (see \cite{D}) for the Dynkin
gradings.

\begin{corollary}
  \label{cor:1.2}
All good $\ZZ$-gradings are among those defined by\break
$\deg
e_{\alpha_i} =-\deg e_{-\alpha_i} = 0$, $1$ or $2$, $i=1,\ldots ,r$.
\end{corollary}


\begin{theorem}
  \label{th:1.3}
Properties (\ref{eq:0.1}) and (\ref{eq:0.2}) of a $\ZZ$-grading
$\fg = \oplus_j \fg_j$ are equivalent.
\end{theorem}

\begin{proof}
  The property $[e,\fg_j] \neq \fg_{j+2}$ for $j \geq -1$ is
  equivalent to the existence of a non-zero $a \in \fg_{-j-2}$
  such that $([e,\fg_j],a)=0$, where $(. \, , \, . )$ is a
  non-degenerate invariant bilinear form on $\fg$.  But the
  latter equality is equivalent to $([e,a], \fg_j)=0$, which is
  equivalent to $a \in \fg^e_{-j-2}$. i.e.,~to $\ad e$ being not
  injective on $\fg_{-j-2}$.
\end{proof}

\begin{theorem}
  \label{th:1.4}
Let $\fg =\oplus_j \fg_j$ be a good $\ZZ$-grading with a good
element $e$.  Then $\fg^e\cong \fg_0 +\fg_{-1}$ as
$\fg^e_0$-modules.
\end{theorem}

\begin{proof}
  Due to (\ref{eq:0.1}) and (\ref{eq:0.2}) we have the following
  exact sequence of $\fg^e_0$-modules:
  \begin{displaymath}
    0 \to \fg^e \to \fg_{-1} + \fg_0 +\fg_+
    \overset{\ad e}{\longrightarrow} \fg_+ \to 0 \, .
  \end{displaymath}

\end{proof}

\begin{corollary}
  \label{cor:1.3}
Let $\fg =\oplus_j \fg_j$ be a $\ZZ$-grading and let $e \in
\fg_2$.  Then $\dim \fg^e \geq \dim \fg_{-1}+\dim \fg_0$, and
equality holds iff $e$ is a good element.

\end{corollary}

\begin{proof}
  We have an exact sequence of vector spaces:
  \begin{displaymath}
    0 \to \fg^{e} \cap (\fg_{-1}+\fg_{\geq}) \to \fg_{-1} +\fg_0 +\fg_+
    \overset{\ad e}{\longrightarrow}
    [e,\fg_{-1}+\fg_{\geq 0}] \to 0 \, .
  \end{displaymath}

Hence $\dim \fg^e + \dim [e,\fg_{-1}+\fg_{\geq}] \geq \dim
\fg_{-1} +\dim \fg_0 +\dim \fg_+$.  Since $\dim
[e,\fg_{-1}+\fg_{\geq}]\leq  \dim \fg_+$ and the equality holds
only if $e$ is good, the corollary follows.

\end{proof}

This result was proved in [FORTW] for even gradings.

\begin{corollary}
  \label{cor:1.4}
Under the conditions of Theorem~\ref{th:1.4}, the representation
of $\fg^e_0$ on $\fg^e$ is self-dual (i.e.,~it is equivalent to
its dual).
\end{corollary}

\begin{proof}
  Consider the bilinear form $\langle a , b \rangle = (e,[a,b])$
  on $\fg_{-1}$.  It is $\fg^e_0$-invariant and, by
  (\ref{eq:0.4}), it is non-degenerate.  Hence the
  $\fg^e_0$-module $\fg_{-1}$ is self-dual.  The $\fg^e_0$-module
  $\fg_0$ is self-dual since the bilinear form $( \, . \, , \,
  . \, ) $ is non-degenerate on $\fg_0$.
\end{proof}

\begin{theorem}
  \label{th:1.5}
Let $\fg = \oplus_j \fg_j$ be a good $\ZZ$-grading with a good
element $e$.  Let $\ft$ be a maximal ad-diagonizable subalgebra
of $\fg^e \cap \fg_0$.  Then all the weights of $\ft$ on $\fg_1$
are non-zero.

\end{theorem}

\begin{proof}
  Let $Z$ be the centralizer of $\ft$ and let $Z'=Z/\ft$.  The
  $\ZZ$-grading of $\fg$ induces a good $\ZZ$-grading
  $Z'=\oplus_j Z'_j$ with a good element $e'$, which is the
  canonical image of $e$.  Note that the centralizer of $e$ in
  $Z'_0$ consists of nilpotent elements, hence by the graded Engel theorem
  \cite{J}, the centralizer of $e'$ in $Z'$ is consists of nilpotent
  elements.  Hence,
  by Corollary~\ref{cor:1.1}, the $\ZZ$-grading of $Z'$ is a
  Dynkin grading.  But it is known that if the centralizer of a
  nilpotent element $e'$ in a reductive Lie algebra $Z'$ consists of
  nilpotent elements, then, for the corresponding Dynkin grading, $Z'_1
  =0$ \cite{C}. This proves the theorem.
\end{proof}

\section{Some examples}
\label{sec:2}

The most popular conjugacy classes of nilpotent elements in a
simple Lie algebra $\fg$ are the following three:~~(a)~the unique
nilpotent orbit of codimension $r(=\rank\, \fg)$, called the
regular nilpotent orbit, (b)~the unique nilpotent orbit of
codimension $r+2$, called the subregular nilpotent orbit, (c)~the
unique (non-zero) nilpotent orbit of minimal dimension, called
the minimal nilpotent orbit.

The characteristic of a regular nilpotent element $e$ is $\Pi =
\Pi_2$.  The reductive part of its centralizer is trivial, hence
by Corollary~\ref{cor:1.1}, the only $\ZZ$-grading for which $e$
is a good element is the Dynkin grading.

Any simple Lie algebra
$\fg$ of rank $r$ has $r$ even $\ZZ$-gradings for which $\dim
\fg_0 =r+2$, corresponding to the characteristics $(j=1, \ldots
,r)$:  $
  \Pi_0 = \Pi \backslash \{ \alpha_j \} \, , \,
  \Pi_2 = \{ \alpha_j \}$.  If $e$ is a good element for such a $\ZZ$-grading, then by
Corollary~\ref{cor:1.3}, $e$ is a subregular nilpotent element.
Furthermore, if $\fg$ is of type different from $A_r$ or $B_r$,
then the reductive part of $\fg^e$ is trivial, hence by
Corollary~\ref{cor:1.1}, only one of the above $r$ even
$\ZZ$-gradings of $\fg$ is good, and it is the Dynkin grading.  We
will show that for the types $A_r$ and $B_r$, all of the above
$r$ even $\ZZ$-gradings are good.

The reductive part of the centralizer of a minimal nilpotent
element $e$ is semisimple for all types except for $A_r$.  Hence,
if $\fg$ is of type different from $A_r$, $e$ is good only for the
Dynkin $\ZZ$-grading.  We will see that for type $A_r$ all good
non-Dynkin
$\ZZ$-grading with a good minimal nilpotent element have the
following characteristics:  $\Pi_2 = \{ \alpha_1 \}$ or $\Pi_2 =
\{ \alpha_r \}$, $\Pi_0 =\Pi \backslash  \Pi_2 $, whereas the
characteristic of the Dynkin grading is: $\Pi_1=\{\alpha_1, \alpha_r \}$,
$\Pi_0=\Pi \backslash \ \Pi_1$.

Recall that $\fg_{\geq}$ is a parabolic subalgebra of $\fg$ and
$\fg_+$ is the nilradical of $\fg_{\geq}$. An element $e$ from the
nilradical of the Lie algebra $\fp$ of a parabolic subgroup
$P\subset G$ is called a \emph{Richardson element} for $\fp$ if the orbit
$Pe$ is open dense in the nilradical of $\fp$.

Note that we have an obvious bijective correspondence between
even $\ZZ$-gradings of $\fg$ and parabolic subalgebras of $\fg$
(by taking $\fg_{\geq}$).

\begin{theorem}
  \label{th:2.1}
Let $\fg = \oplus_{j \in \ZZ} \fg_{2j}$ be an even $\ZZ$-grading,
and let $\fg_{\geq}$ be the corresponding parabolic subalgebra of
$\fg$.  Then $e \in \fg_2$ is a Richardson element for
$\fg_{\geq}$ iff $e $ is good.
\end{theorem}

\begin{proof}
  It is well known that $\fg^e \subset \fp$ for a Richardson
  element $e$ of the parabolic subalgebra $\fp$.  Hence if a
  Richardson element $e$ lies in the $\fg_2$ then the
  $\ZZ$-grading is good.

Conversely, if $e$ is good, then, by Theorem~\ref{th:1.4}, $\dim
\fg^e =\dim \fg_0$.  Hence, since $\fg^e \subset \fg_{\geq}$, we
obtain that $\dim \fg_{\geq} (e)=\dim \fg_{\geq}-\dim \fg_0 =
\dim \fg_+$.  Hence $e$ is a Richardson element of the parabolic
subalgebra $\fg_{\geq}$.
\end{proof}

A Richardson element defined by Theorem~\ref{th:2.1} is called
\emph{good}.

\section {Nilpotent orbits and parabolic subalgebras in classical
  Lie  algebras}
\label{sec:3}

Let $V$ be a vector space over $\FF$ of finite dimension $n$. Denote
by $\mathrm{Gl}_n$ the group of automorphisms of $V$ and by
$\fg\fl_n$ its Lie algebra.

Let $Par(n)$ be the set of \emph{partitions of $n$}, i.~e. the set
of all tuples $p=(p_1,\ldots,p_s)$ with $p_i\in\NN$, $p_i \geq
p_{i+1}$ and $p_1+\cdots+p_s=n$. It will be convenient to assume that
in fact $p$ has arbitrary number of further components on the
right, all equal to zero, i.~e. $p_{s+1}=p_{s+2}=\cdots=0$. We will
denote by $\mathrm{mult}_p(j)$ multiplicity of the number $j$ in
the partition $p$, i.~e.
$$
\mathrm{mult}_p(j)=\#\{i:p_i=j\},
$$
so that the partition $p$ can be also written as
\begin{equation}\label{pmult}
\sum_{i\geq1}i\mathrm{mult}_p(i)=n.
\end{equation}
For a partition $p=(p_1,\ldots,p_s) \in Par(n)$ or, more generally, for
any sequence (not necessarily decreasing, but such that only
finitely many of its entries are positive) we denote by $p^*$ the
dual partition, so $p^*=(p^*_1,p^*_2,\ldots)$ with $p^*_j:=\#\{i:p_i
\geq j\}$, $j=1,2,\ldots$. Note in particular that
\begin{equation}\label{mult}
\mathrm{mult}_p(j)=p^*_j-p^*_{j+1}.
\end{equation}

The nilpotent conjugacy classes in $\fg\fl_n$ correspond
bijectively to the partitions of $n$ (Jordan normal form). So we
denote a conjugacy class in $\mathrm{Gl}_n$ which corresponds to
the partition $p$, by $\mathrm{Gl}_n(p)$.  The following is well known (see \cite {C}):
\begin{equation}\label{3.3}
\dim\mathrm{Gl}_n(p)=n^2-(( p^*_1)^2+( p^*_2)^2+\cdots)\, .
\end{equation}
The simple Lie algebra $\fs\fl_n$ (of type $A_{n-1}$) is the
subalgebra of $\fg\fl_n$ consisting of traceless $n\times n$
matrices over $\FF$. We will take as its Cartan subalgebra $\fh$ its
subspace of traceless diagonal matrices. The roots and weights
live in the dual $\fh^*$ of $\fh$, which can be identified with
the subspace $x_1+\cdots+x_n=0$ of $\RR^n$. The roots are $\{e_i -e_j
| 1\leq i\neq j \leq n \}$, and we will choose the positive ones to
be $\Delta_+=\{e_i -e_j | 1\leq i<j \leq n \}$. The simple roots
are then $\alpha_i =e_i -e_{i+1}$, for $1\leq i \leq n-1$.
It is obvious that the description of nilpotent elements,
parabolic subalgebras, as well as good gradings, etc, for $\fg\fl_n$
is the same as for $\fs\fl_n$.

It is well known that there exists a bijection between the
conjugacy classes of parabolic subalgebras of $\fg\fl_n$ (and $\fs\fl_n$) and
compositions of $n$. An ordered sequence of positive integers
$(a_1,\ldots,a_m)$ is called a composition of $n=\sum a_i$. Under
the above bijection, a parabolic subalgebra corresponding to the
composition $(a_1,\ldots,a_m)$ is the one consisting of elements
preserving a flag $\{0\}= V_0 \subset V_1\subset\cdots\subset V_{m-1}\subset
V_m=V$, where $\dim V_i/V_{i-1}=a_i$, $i=1,\ldots ,m$.



Consider a $2n$- (resp.~$2n+1$)-dimensional vector space $V$ over
$\FF$ with a basis $v_1,\ldots,v_n,
v_{-1}, \ldots ,v_{-n}$
(resp.~$v_1,\ldots ,v_n, v_0,v_{-1},\ldots,v_{-n}$).  In the
case of $\dim V=2n$, denote by $\langle \, . \, , \, . \, \rangle$ the
skew-symmetric bilinear form on $V$ defined by $\langle v_i,v_{-j}
\rangle =\delta_{ij}$ for $i \geq 1$.  Denote by  $( \, . \, , \,
. \, )$ the symmetric bilinear form on $V$ defined by
$(v_i,v_{-j})=\delta_{ij}$.  Let $G =Sp_{2n} =\{ g \in GL_{2n} |
\langle gu,gv \rangle = \langle u,v \rangle ,u,v \in V \}$ be the
corresponding symplectic group and $\fg =\fs\fp_{2n}$ its Lie
algebra, and let $G=O_N=\{ g \in GL_N | (gu,gv ) = (u,v),u,v \in
V \}$, where $N =2n$ or $2n+1$, be the corresponding orthogonal
group and $\fg =\fso_N$ its Lie algebra.

The parabolic subalgebras of $\fg =\fsp_N$ or $\fso_N$ are
described as stabilizers of \emph{isotropic} flags in $V$,
i.e.,~the flags $V_0 =\{ 0 \} \subset V_1 \subset \cdots \subset V_m =V$ such that
$V^{\perp}_j = V_{m-j}$ for $j=0,1,\ldots ,m$.  We let $a_i =\dim
V_i /V_{i-1}$, $i \geq 1$, let $t=[m/2] $ and $q=0$ if $m$ is
even, and $=a_{t+1}$ if $m$ is odd.  Then $(a_1 ,\ldots ,a_t)$ is
a composition of $\tfrac{1}{2}(N-q)$.  Thus we get a bijective
correspondence between conjugacy classes of parabolic subalgebras
of $\fg$ and the pairs $(a_1 ,\ldots ,a_t \, ; \, q)$, where $q$
is an integer between $0$ and $N$ such that $N-q$ is even, and
$(a_1,\ldots ,a_t)$ is a composition of $\tfrac{1}{2}(N-q)$;
also, in the case $\fso_{2n}$, $q \neq 2$.

Denote by $SPar(N)$ (resp.~$OPar(N)$ the subset of $Par(N)$
consisting of partitions whose odd (resp.~even) parts occur with even
multiplicity.
It is well known that the nilpotent orbits of $G$ in $\fg
=\fs\fp_N$ and $\fs\fo_N$
correspond bijectively to the partitions from $SPar(N)$ and $OPar(N)$
respectively (see \cite{C}).

A description of the reductive parts of centralizers $\fg^e$ for
nilpotent elements $e$ in Lie algebras of classical type can be
found in \cite{C}. From this description the following theorem easily follows.

\begin{theorem}
  \label{th:3.1}
  Let $\fg$ be a simple Lie algebra of classical type, let $e=e(p)$
be the nilpotent element corresponding to a partition $p$, and let
$c(e)$ be the dimension of the center of $\fg^\fs$, the reductive part of the
centralizer $\fg^e$(see Lemma~\ref{lem:1.2}). Then

  a) $c(p)=\#($different parts of the partition $p)-1$ in the
  case $\fg=\fs\fl_n$;

  b) $c(p)=\#($odd parts of $p$ with multiplicity 2$)$ in the
  case $\fg=\fs\fo_n$;

  c) $c(p)=\#($even parts of $p$ with multiplicity 2$)$ in the
  case $\fg=\fs\fp_n$.
\end{theorem}

\section{Good gradings of $\fg\fl_n$}
\label{sec:4}

A \emph{pyramid} $P$ is a finite collection of boxes of size
$1 \times 1$ on the plane, centered at $(i,j)$, where $i,j \in \ZZ$, such that
the second coordinates of the centers of boxes of the lowest row equal 1,
the first coordinates of $j^{th}$ row form
an arithmetic progression $f_j, f_j + 2,\ldots, l_j$ with difference 2,
$f_1 = -l_1$, and one has
$$
\begin{matrix}
f_j&\leq &f_{j+1} , l_j&\geq &l_{j+1} &\textrm{for all } j.
\end{matrix}
\leqno{(4.1)}
$$
The \emph{ size } of a pyramid is the number of boxes in it.

To a given pyramid $P$ of size $n$ we associate a nilpotent
endomorphism of the vector space $\FF^n$ in the
following manner: enumerate the squares of $P$ in some order,
label the standard
basis vectors of $\FF^n$ by the boxes with the corresponding
number, and define an endomorphism $e(P)$ of $\FF^n$ by letting it
act ``along the rows of the pyramid'', i.~e., by sending the
basis vector labeled by a box to the basis vector labeled by
its neighbor on the right, if this neighbor belongs to $P$ and to
$0$ otherwise. Denote by $p_j$ the number of squares on the $j$-th
row of $P$. Then the endomorphism $e(P)$ is a
nilpotent one corresponding to the partition (i.~e. with sizes of
Jordan blocks given by) $p=(p_1,\ldots,p_k)$ and the endomorphisms
corresponding to all pyramids with $n$ boxes and fixed
lengths of rows belong to the conjugacy class of the nilpotent
$e(P)$. Define a diagonal matrix $h(P)\in \fg\fl_n$ by letting its $j^{th}$
diagonal entry equal to the first coordinate of the center of the $j^{th}$
box. Then the eigenspace decomposition of $ad(h(P))$ is a $\ZZ$-grading
of $\fg\fl_n$. The characteristic of this $\ZZ$-grading can be described as
follows.
First, denote by $h_j$ the number of squares in
in the $j$-th column of $P$, $j=1,\ldots,2p_j -1$,
i.~e. such that the first coordinate of
their centers is equal to $j$. Note that for $j$ odd, necessarily
$h_j>0$.
Next, for each $h_j$ construct a sequence
which begins with $h_j -1$ zeros and is followed by either
$2$ --- if the right neighbor of $h_j$ is zero; or by $1$ --- if
the right neighbor of $h_j$ is nonzero; or by nothing at all, if
$h_j$ does not have any right neighbors (i.~e. $j=2p_1-1$).
Then concatenate the sequences obtained, to form the sequence
of $n-1$ integers equal $0$, $1$ or $2$,
which defines the characteristic in question by assigning
these integers to the corresponding simple roots.
It is also easy to see that an elementary matrix $E_{i,j}$ has a non-negative
degree in this grading iff the label
$i$ is not located strictly to the left of the label $j$ in the
pyramid $P$.

Given a partition $p=(p_1,\ldots,p_k) \in Par(n)$,
denote by $P(p)$ the symmetric pyramid corresponding to $p$,
i.e. the pyramid with $k$ rows such that the $j^{th}$ row
contains $p_j$ boxes centered at $(i,j)$, where $i$ runs over
the arithmetic progression with difference 2 and $f_j=-p_j
+1 =-l_j$. Denote by $\Pyr(p)$ the set of all pyramids
attached to the partition $p$, i.e. the pyramids containing $p_j$
boxes in the $j^{th}$ row, $j=1,\ldots,k$.
Obviously, all the pyramids from $\Pyr(p)$ are obtained from the symmetric
one, $P(p)$, by a horizontal shift for each $j>1$ of the boxes of the $j^{th}$
row (as a whole)
in such a way that condition (4.1) is satisfied. Hence we
obtain the following lemma.
\begin{lemma}
\label{lem:4.1}
$$
\#Pyr(p)=\prod_{i=1}^{k-1}(2(p_i -p_{i+1})+1)
$$
(here in the case $k=1$, by the empty product is understood  to be 1, as
usual).
\end{lemma}

Using this lemma, we can calculate the generating function
$$
F(q)=\sum_n\mathrm{Pyr}_n q^n
$$
for the numbers $\Pyr_n$ of pyramids of size $n$. Indeed, using
notation from the lemma we obviously have
$$
\mathrm{Pyr}_n=\sum_{p\in Par(n)}\#Pyr(p).
$$
Thus according to the lemma, we can write
$$
F(q)=\sum_p\left(\prod_{i:p_{i+1}>0}(2(p_i
-p_{i+1})+1)\right)q^{\sum_ip_i},
$$
with the sum ranging over all partitions of all natural numbers.
Observe now that since partitions are in one-to-one correspondence
with duals of partitions, we obviously have
$$
F(q)=\sum_p\left(\prod_{i:p^*_{i+1}>0}(2(p^*_i
-p^*_{i+1})+1)\right)q^{\sum_ip_i}.
$$
Then using \ref{pmult} and \ref{mult} we can write
$$
F(q)=\sum_p\left(\prod_{i:p^*_{i+1}>0}(2\mathrm{mult}_p(i)+1)\right)
q^{\sum_{i\geq1}i\mathrm{mult}_p(i)}.
$$
Now observe that for any $i$, the condition $p^*_{i+1}>0$, i.~e.
$\#\{j:p_j\geq i+1\}>0$, is equivalent to $i<p_1$. Thus the above
can be rewritten as
$$
F(q)=\sum_p \left( \prod_{i<p_1} (2\mathrm{mult}_p(i)+1)
  q^{i\mathrm{mult}_p(i)} \right) q^{p_1\mathrm{mult}_p(p_1)},
$$
or as well
$$
F(q)=\sum_n \sum_{p:p_1=n} \left(\prod_{i<n}
  (2\mathrm{mult}_p(i)+1) q^{i\mathrm{mult}_p(i)} \right)
  q^{n\mathrm{mult}_p(n)}.
$$
The last expression can be rewritten as
\begin{eqnarray*}
\lefteqn{\sum_n \sum_{m_1,m_2,\ldots,m_{n-1} \geq0, m_n>0}
    \left( (2m_1+1) q^{m_1}
     \cdots
     (2m_{n-1}+1)q^{{n-1}m_{n-1}}
     \right)q^{nm_n}}\\
  &=& \!\!\!
    \sum_n \left( \sum_{m_1\geq0} (2m_1+1)q^{m_1}
    \cdots \!\!\!\!
    \sum_{m_{n-1}\geq0}(2m_{n-1}+1)(q^{n-1})^{m_{n-1}}
    \right) \! \sum_{m_n>0}(q^n)^{m_n}.
\end{eqnarray*}
Taking into account that (for $|t| < 1$)
$$
\sum_{m\geq0}(2m+1)t^m=\frac{1+t}{(1-t)^2},
\sum_{m>0}t^m=\frac t{1-t},
$$
we obtain the following
\begin{proposition}
\label{genf}
\label{prop:4.1}
The generating function for the numbers $\mathrm{Pyr}_n$ is given
by
$$
F(q)=\sum_{n\geq1} \left(\prod_{k=1}^{n-1} \frac{1+q^k}
  {(1-q^k)^2}\right) \frac{q^n}{1-q^n}.
$$
\end{proposition}

\begin{theorem}
\label{th}
\label{th:4.0}
 Let $p$ be a partition of the number $n$, $P$
a pyramid from $\Pyr(p)$ and $h(P)$ the corresponding diagonal
matrix in $\fg\fl_n$. Then the
pair $(h(P),e(p))$ is good.
\end{theorem}
\begin{proof}
  First of all, we recall that $e(p)$ is the endomorphism which
  acts ``along the rows of the pyramid'' and for this reason it
  is natural to depict it via horizontal arrows which connect
  centers of boxes with their right neighbors on the same row.
  Endomorphisms $E_{i,j}$ map the $j$th basis vector of $\FF^n$
  to the $i$th basis vector, and we represent $E_{i,j}$ by the
  arrow, which connects the corresponding centers of the boxes of
  the pyramid $P$. Then endomorphisms commuting with $e(p)$ are
  precisely those represented by collections of arrows, which fit
  with arrows of $e(p)$ into commutative diagrams.
  Figures~\ref{fig:1}, \ref{fig:2}, \ref{fig:3} represent
  examples of such commutative diagrams. The loops in these
  pictures mean identity mappings. Since the end of no arrow in
  these diagrams is located strictly to the left of its source,
  all corresponding endomorphisms have non-negative degree with
  respect to $h(P)$. It is easy to see that when $i$ runs through
  the set $[1,\ldots,k]$, the corresponding endomorphism from
  figures~\ref{fig:1}, \ref{fig:2}, \ref{fig:3} are linearly
  independent. The number of the first (resp.~2nd and 3d) type
  diagrams is $p_1+p_2+\cdots+p_k$
  (resp.~$2\sum_{i=1}^{k}(i-1)p_i)$.  It is well known (see
  \cite{C}) that $n+2\sum_{i=1}^{i=k}(i-1)p_i=(p_1^*)^2+ \cdots
  +(p_k^*)^2$. All this means that diagrams of types 1, 2, 3,
  form a basis of the centralizer of the nilpotent $e(p)$ and all
  elements of this centralizer have non-negative degree with
  respect to the $\ZZ$-grading determined by $h(P)$.
\end{proof}

\begin{figure}[tbp]
  \begin{center}
    $$
\begin{array}{c}
\xy
0;<5.8mm,0mm>:<0mm,5mm>::
(0,0)*{\circ}="f",(0,-1)*{_{f_i}},
(1.75,0)*{\circ}="f+1",(1.75,-1)*{_{f_i+1}},
(3.75,0)*{\dots},
(5.75,0)*{\circ}="f+r",(5.75,-1)*{_{f_i+r}},
(7.5,0)*{\circ}="f+r+1",(7.5,-1)*{_{f_i+r+1}},
(10,0)*{\dots},
(12.5,0)*{\circ}="l-r-1",(12.5,-1)*{_{l_i-r-1}},
(14.25,0)*{\circ}="l-r",(14.25,-1)*{_{l_i-r}},
(16.25,0)*{\dots},
(18.25,0)*{\circ}="l-1",(18.25,-1)*{_{l_i-1}},
(20,0)*{\circ}="l",(20,-1)*{_{l_i}},
"f"\ar"f+1"\POS"f+1"\ar@{-}(2.75,0)
\POS(4.75,0)\ar"f+r"\POS"f+r"\ar"f+r+1"\ar@{-} (6.75,1.1)
\POS"f+r+1"\ar@{-} (8.5,0) \ar@{-}(8.5,1.1)
\POS(11.5,1.1)\ar"l-r-1"
\POS(11.5,0)\ar"l-r-1"\POS"l-r-1"\ar"l-r"\POS(13.25,1.1) \ar"l-r"
\POS"l-r" \ar@{-}(15.25,0)
\POS(17.25,0)\ar"l-1"\POS"l-1"\ar"l"
\POS"f"\ar@/^5ex/"f+r"
\POS"f+1"\ar@/^5ex/"f+r+1"
\POS"l-r-1"\ar@/^5ex/"l-1"
\POS"l-r"\ar@/^5ex/"l"
\endxy 
\end{array}
$$
    \caption{}
    \label{fig:1}
  \end{center}
\end{figure}

\begin{figure}[tbp]
  \begin{center}
    $$
\begin{array}{c}
\xy
0;<10mm,0mm>:<0mm,3mm>::
(-2,0)*{(i)},
(-2,5)*{(j)},
(0,0)*{\circ}="fi",(0,-1)*{_{f_i}},
(2,0)*{\circ}="fi+1",(2,-1)*{_{f_i+1}},
(3,0)*{\dots},
(4,0)*{\circ}="fi+lj-fj-r-1",(4,-1)*{_{f_i+l_j-f_j-r-1}},
(6,0)*{\circ}="fi+lj-fj-r",(6,-1)*{_{f_i+l_j-f_j-r}},
(8,0)*{\dots},
(9.75,0)*{\circ}="li",(9.75,-1)*{_{l_i}},
(1,5)*{\circ}="fj",(1,6)*{^{f_j}},
(2,5)*{\dots},
(3,5)*{\circ}="fj+r",(3,6)*{^{f_j+r}},
(5,5)*{\circ}="fj+r+1",(5,6)*{^{f_j+r+1}},
(6,5)*{\dots},
(7,5)*{\circ}="lj-1",(7,6)*{^{l_j-1}},
(9,5)*{\circ}="lj",(9,6)*{^{l_j}},
"fi"\ar"fi+1"\POS"fi+1"\ar@{-}(2.5,0)
\POS(3.5,0) \ar"fi+lj-fj-r-1" \POS"fi+lj-fj-r-1" \ar"fi+lj-fj-r"
\POS"fi+lj-fj-r" \ar@{-}(6.5,0) \POS(9.25,0)\ar"li"
\POS"fj"\ar@{-}(1.5,5) \POS(2.5,5) \ar"fj+r" \POS"fj+r"
\ar"fj+r+1" \POS"fj+r+1" \ar@{-}(5.5,5) \POS(6.5,5) \ar"lj-1"
\POS"lj-1" \ar"lj"
\POS"fi"\ar"fj+r"
\POS"fi+1"\ar"fj+r+1"
\POS"fi+lj-fj-r-1"\ar"lj-1"
\POS"fi+lj-fj-r"\ar"lj"
\endxy 
\end{array}
$$
    \caption{}
    \label{fig:2}
  \end{center}
\end{figure}

\begin{figure}[tbp]
  \begin{center}
    $$
\begin{array}{c}
\xy
0;<10mm,0mm>:<0mm,3mm>::
(-2,0)*{(i)},
(-2,5)*{(j)},
(0,5)*{\circ}="fj",(0,6)*{^{f_j}},
(2,5)*{\circ}="fj+1",(2,6)*{^{f_j+1}},
(3,5)*{\dots},
(4,5)*{\circ}="lj-r-1",(4,6)*{^{l_j-r-1}},
(6,5)*{\circ}="lj-r",(6,6)*{^{l_j-r}},
(7,5)*{\dots},
(8,5)*{\circ}="lj",(8,6)*{^{l_j}},
(-1,0)*{\circ}="fi",(-1,-1)*{_{f_i}},
(1,0)*{\dots},
(3,0)*{\circ}="li+fj-lj+r",(3,-1)*{_{l_i+f_j-l_j+r}},
(5,0)*{\circ}="li+fj-lj+r+1",(5,-1)*{_{l_i+f_j-l_j+r+1}},
(6,0)*{\dots},
(7,0)*{\circ}="li-1",(7,-1)*{_{l_i-1}},
(9,0)*{\circ}="li",(9,-1)*{_{l_i}},
"fj"\ar"fj+1"\POS"fj+1"\ar@{-}(2.5,5)
\POS(3.5,5) \ar"lj-r-1" \POS"lj-r-1" \ar"lj-r" \POS"lj-r"
\ar@{-}(6.5,5) \POS(7.5,5)\ar"lj"
\POS"fi"\ar@{-}(-.5,0) \POS(2.5,0) \ar"li+fj-lj+r"
\POS"li+fj-lj+r" \ar"li+fj-lj+r+1" \POS"li+fj-lj+r+1"
\ar@{-}(5.5,0) \POS(6.5,0)\ar"li-1"\POS"li-1"\ar"li"
\POS"fj"\ar"li+fj-lj+r"
\POS"fj+1"\ar"li+fj-lj+r+1"
\POS"lj-r-1"\ar"li-1"
\POS"lj-r"\ar"li"
\endxy 
\end{array}
$$
    \caption{ }
    \label{fig:3}
  \end{center}
\end{figure}

\begin{theorem}
  \label{th:4.1}
Let $e(p)$ be the nilpotent element defined by a partition
$$p=(p_1^{m_1},p_2^{m_2},\ldots,p_d^{m_d})=
(\underbrace{p_1,\ldots,p_1}_{m_1},\underbrace{p_2,\ldots,p_2}_{m_2},\ldots,
\underbrace{p_d,\ldots,p_d}_{m_d})$$ of the number $n$. Define
$t_i:=\sum_{j=1}^{j=i}m_jp_j$ for $1\leq i\leq d$, so that
$t_1<t_2< \cdots <t_d=n$. Let $P(p)$ be the symmetric
pyramid determined by the partition $p$, let $h(p):=h(P(p))$ be the
corresponding diagonal matrix in $\fg\fl_n$ and let
$h=(h_1,h_2,\ldots,h_n)$ be a
diagonal matrix. Then, the pair $(h(p)+h,e(p))$ is good if and
only if the coordinates $h_i$ satisfy the following conditions:
\begin{enumerate}
\item $h_i-h_j$ are integers,
\item $h_1=h_2= \cdots =h_{t_1}, h_{t_1+1}= \cdots =h_{t_2},
  \ldots ,
  h_{t_{d-1}+1}= \cdots =h_{t_d}$,
\item $|h_{t_1}-h_{t_2}|\leq p_1-p_2,  \ldots, | h_{t_{d-1}}-h_{t_d}|\leq p_{d-1}-p_d$,
\item $\sum_{i=1}^{n}h_i=0$.
\end{enumerate}
Moreover if for each $i\in[1,d-1]$ we set
$h_{t_{i-1}}-h_{t_i}=a_i$, where
$a_i\in[-p_{i-1}+p_i,p_{i-1}-p_i]$, then the system of linear
equations (conditions 2 and 4) has
 a unique solution.
\end{theorem}
\begin{proof}
  Identify the boxes of the first $m_1$ rows (resp.~the second
  $m_2$ rows), \ldots , (resp.~the last $m_d$ rows) of the
  symmetric pyramid $P(p)$ with the first $t_1$ (resp.~the second
  $t_2$), \ldots, (resp.~the last $t_d$) basis vectors of the
  base space. Then, because the number of boxes in first $m_1$
  rows is equal to $p_1$ (resp.~that in the second $m_2$ rows is
  equal to $p_2$), \ldots , (resp.~that in the last $m_d$ rows is
  equal $p_d$), it follows that for $h$ to be contained in the
  center of the reductive part of the centralizer, it is
  necessary and sufficient that the condition 2 be satisfied.
  Theorem~\ref{th:4.1} gives a transparent description of the
  basis of the centralizer of $e(p)$. This description together
  with equalities $E_{i,j}=[E_{i,k},E_{k,j}]$;
  $[h(p),e(p)]=2e(p)$ and condition 2, tell us that
  non-negativity of $h$ follows from non-negativity of $h$ at the
  extreme elements of the centralizer, i.e.~the elements
  corresponding to the diagrams in Figures~\ref{fig:2} or
  \ref{fig:3} with $t_i<t_{i+1}$ and $r=0$. The inequalities 3
  are just equivalent to this non-negativity. The last statement
  of the theorem is clear because the determinant of the
  corresponding system of linear equations is~$n$.
\end{proof}

In the notation of Theorem~\ref{th:4.1} put $a_i=1$ and $a_j=0$ for
$j\neq i$. Then solving the corresponding system of linear
equations we obtain $h_1= \cdots =h_{t_{i-1}}=\frac{m_i+ \cdots +m_d}n$ and
$h_{t_{i-1}+1}= \cdots =h_n=1-h_1$. The grading on $\fs\fl_n$
determined by $h+h(p)$ will not change if we subtract from
$h+h(p)$ the scalar matrix $h_1I_n$. This semisimple element, as
is not difficult to see, corresponds to the grading determined by
the pyramid obtained from the symmetric pyramid $P(p)$ by
shifting to the left by 1 as one whole all rows starting from
$t_{i-1}+1$ up.

Thus, Theorem~\ref{th:4.1} produces a one-to-one correspondence between the
pyramids from $\Pyr(p)$ and good gradings for the nilpotent
$e(p)$. Consequently  Lemma~\ref{lem:4.1}
gives the number of good pairs of the form $(h+h(p),e(p))$
and the generating function for the number of good gradings of
$\fs\fl_n$.

Let us recall that a $\ZZ$-grading is called even if dim$\fg_i=0$
for $i\equiv1 (\mod2)$. In the $\fs\fl(n)$ case a partition $p$
determines an even grading $h(p)$ if and only if all parts $p_i$
of $p$ have the same parity. In terms of the pyramids
$P\in$Pyr$(p)$ this means that the first coordinates of the centers of
all boxes constituting the pyramid $P$ have the same parity.

\begin{proposition}
\label{prop:4.2}
Let $e(p)$ be a nilpotent element of $\fs\fl_n$ determined by a
partition $p$. Then there exists a semisimple element $h^e$ such
that the pair $(h^e,e(p))$ is good and $h^e$ determines an even
grading.
\end{proposition}

\begin{proof}
If the nilpotent $e(p)$ is even, then one can take $h^e=h(p)$,
the semisimple element determining the Dynkin grading.

Let $i_1$, \ldots , $i_k\in\{1, \ldots ,d\}$ be all those natural numbers
$i$ for which $p_{i-1}-p_i\equiv 1 (\mod2)$. Put
$a_{i_1}=a_{i_2}= \cdots =a_{i_k}=1$, $a_j=0$ for $j\ne i_1, \ldots ,i_k$
(see the notation of Theorem~\ref{th:4.1}) and denote by
$h(a_{i_1}, \ldots ,a_{i_k})$ the solution of the corresponding system
of equations. Then $h^e:=h(a_{i_1}, \ldots ,a_{i_k})+h(p)$ will be the
required semisimple element.
\end{proof}

Recall that a \emph{unimodal sequence} of size
$n$ is a sequence of natural numbers $h_1\le h_2\le \cdots \le h_i\geq
h_{i+1}\geq \cdots \geq h_k$ satisfying $\sum_{j=1}^l h_j=n$.

\begin{proposition}
\label{prop:4.3}
There exists a one-to-one correspondence between even good
gradings of the simple Lie algebra $\fs\fl_n$ and unimodal sequences
of size $n$.
\end{proposition}
\begin{proof}
Given a unimodal sequence $h=(h_1, \ldots ,h_k)$ of size $n$, we
construct a pyramid in the following way: the first
row of the pyramid will consist of $k$ boxes, with the first
coordinates constituting an arithmetic progression with difference
$2$ and first entry $-k+1$. The pyramid will consist of $2k-1$
columns, with columns at even places consisting of 0 boxes and the
column at the $(2i-1)$\st{th} place consisting of $h_i$ boxes,
$i=1,2,\ldots ,k$. This
pyramid belongs to the set Pyr$(h^*)$ and determines an even good
grading.

Conversely the sequence of nonzero column heights of a
pyramid~$P$ determined by an even grading will be an unimodal
sequence of size~$n$. It is clear that these two mappings are
mutually converse.
\end{proof}

\begin{corollary}
\label{cor:4.1}
The generating function for the numbers of even good gradings of
the Lie algebras $\fs\fl_n$, $n=1,2, \ldots$, is
$$
U(q)=\sum_{n\geq1}(-1)^{n+1}q^{\binom{n+1}2}\prod_{k\geq1}\frac1{(1-q^k)^2}
\eqno{(*)}
$$
\end{corollary}

\begin{proof}
This follows directly from the previous proposition since
according to (\cite{S}, Corollary 2.5.3), the generating function
for unimodal sequences is $U(q)$.
\end{proof}

\begin{remark}
\label{rem:4.1}
If one looks at the proof of the aforementioned Corollary 2.5.3
from \cite{S}, there the generating function is obtained by
transforming in a clever way the series
$$
\sum_{n\geq1}\left(\prod_{k=1}^{n-1}\frac1{(1-q^k)^2}\right)\frac{q^n}{1-q^n},
$$
which resembles our generating function $F(q)$ from
Proposition \ref{genf}. This makes one wonder whether the latter
can be similarly transformed to a more satisfactory form. Now the
analog of the second factor of (*) for the series $F(q)$ is more
or less obviously the product
$$
\prod_{k\geq1}\frac{1+q^k}{(1-q^k)^2},
$$
so a natural thing to do is to look at the result of dividing
$F(q)$ by this product. The result gives the equality
$$
F(q)=\sum_{n\geq1} \left( \prod_{k=1}^{n-1}
  \frac{1+q^k}{(1-q^k)^2} \right) \frac{q^n}{1-q^n}=
\sum_{n\geq1} (q^{\frac{3n^2-n}2}-q^{\frac{3n^2+n}2})
\prod_{k\geq1} \frac{1+q^k}{(1-q^k)^2}.
$$
We are grateful to G. Andrews for providing a proof of this identity.
\end{remark}

\begin{corollary}
\label{cor:4.2}
Let $\fp(a_1, \ldots ,a_m)$ be a parabolic subalgebra of $\fs\fl_n$
corresponding to the composition $(a_1, \ldots ,a_m)$.  Then a
Richardson element of
$\fp(a_1, \ldots ,a_m)$ is good for the corresponding even
$\ZZ$-grading of $\fsl_n$ iff the
sequence $(a_1, \ldots ,a_m)$ is unimodal.
\end{corollary}

This result and some necessary conditions for other classical groups
were obtained by Lynch \cite{L}.

\section{Good gradings of $\fsp_{2n}$}
\label{sec:5}

The proofs of the results of this and the next section are
similar to those for $\fs\fl_n$, and will be omitted.

{}From now on, given a partition $p$, we denote by $p_1 >p_2
>\cdots$ its distinct non-zero parts and use notation
$p=(p^{m_1}_1, \ldots , p^{m_s}_s)$, where $m_i$ is the
multiplicity of $p_i$ in $p$.  A partition is called
\emph{symplectic} (resp.~\emph{orthogonal}) if $m_i$ is even for
odd (resp.~even) $p_i$.  Recall that symplectic partitions of $2n$ correspond
bijectively to nilpotent orbits in $\fsp_{2n}$.

Given a symplectic partition $p$ of $2n$, construct a symplectic
pyramid $SP (p)$ as follows.  It is a centrally symmetric (around
$(0,0)$) collection of $2n$ boxes of size~$1\times 1$ on the plane,
centered at points with integer coordinates (called the
coordinates of the corresponding boxes).  The $0$\st{th} row of
$SP(p)$ is non-empty iff $m_1=2k_1 +1$ is odd, and in this case
the first coordinates of boxes in this row form an arithmetic
progression $-p_1 +1, -p_1+3, \ldots ,p_1-1$.  The rows from
1\st{st} to $k$\st{th} consist of boxes with the first
coordinates forming the same arithmetic progression.  If
the multiplicity $m_2$ of $p_2$ is
even, then the rows from $k_1$\st{th}$+1$ to $\left(
k_1+\tfrac{m_2}{2}\right)$\st{th} consist of boxes with first
coordinates forming the arithmetic progression $-p_2 +1,
-p_2+3,\ldots , p_2-1, $.  However, if $m_2$ is odd, then the
$k_1+1$\st{th} row consists of boxes with first coordinates
forming the arithmetic progression $1,3,\ldots ,p_2-1$ (recall
that $p_2$ must be even if $m_2$ is odd) and the $m_2-1$
subsequent rows consist of boxes with first coordinates forming
the arithmetic progression $-p_2 +1, -p_2+3 ,\ldots , p_2-1$.
All the subsequent parts of $p$ are treated in the same way as $p_2$.
The rows in the lower half-plane are obtained by the
central symmetry.

The nilpotent symplectic endomorphism $e(p)$ of $\FF^{2n}$
corresponding to a symplectic partition $p$ is obtained by
filling the boxes of $SP (p)$ by vectors $v_1 , \ldots ,v_n ,
v_{-1}, \ldots ,v_{-n}$ such that vectors in boxes in the right
half-plane ($x \geq 0$ and $y>0$ if $x=0$) have positive
indices~$i$ and those in the centrally symmetric boxes have
indices~$-i$.  Then $e(p)$ maps vectors in each box to its right
neighbor, and to $0$ if there is no right neighbor, with the
exception of the boxes with coordinates $(-1,-\ell)$ and no right
neighbors; the vector in such a box is mapped by $e(p)$ to the
vector in the $(1,\ell)$ box (which has no left neighbors).

It is easy to see that the eigenvalue on a vector $v_i$ of the
diagonal matrix $h(p) \in \fsp_{2n}$, which defines the
corresponding to $e(p)$ Dynkin grading, is equal to the first
coordinate of the corresponding box of $SP(p)$.

Recall that, by Theorem~\ref{th:3.1}, the dimension of the center
of the reductive part of $\fsp^{e(p)}_{2n}$ is equal to $c(p)$,
the number of even parts of the partition $p$ having
multiplicity~$2$.  If $c(p) =0$, then, by
Corollary~\ref{cor:1.1}, the only good grading of $\fs\fp_{2n}$
with good element $e(p)$ is the Dynkin one.  Thus, we may assume
from now on that $c(p)>0$.

\begin{lemma}
  \label{lem:5.1}
Let $p_1 ,\ldots ,p_{c(p)}$ be all distinct even parts of a
symplectic partition $p$, having multiplicity 2.
Define diagonal matrices $z (t_1 ,\ldots , t_{c(p)}) \in
\fsp_{2n}$, $t_1 , \ldots ,t_{c(p)} \in \FF$, whose $i$\st{th}
diagonal entry is $t_i$ if the basis vector lies in a box of
$SP(p)$ in the (strictly) upper
half-plane in a row corresponding to the part $p_i$, and is
$-t_i$ if the basis vector lies in the centrally symmetric box, and all
other entries are zero.  Then the center of the reductive part of
$\fsp^{e(p)}_{2n}$ consists of all these matrices.
\end{lemma}

\begin{theorem}
  \label{th:5.1}
In notation of Lemma~\ref{lem:5.1}, the element $H(p \, ; \,t_1,\ldots
,t_{c(p)})\break
:=h(p)+z(t_1 ,\ldots ,t_{c(p)})$ defines a good
$\ZZ$-grading of $\fsp_{2n}$ iff one of the following cases holds:

\romanparenlist
\begin{enumerate}
\item 
all parts of $p$ are even and have multiplicity $2$, and either
all $t_i \in \{ -1,0,1 \}$ or all $t_i \in \{ \tfrac{1}{2}, -\tfrac{1}{2}\}$;

\item 
not all parts of $p$ are even of multiplicity $2$, and all $t_i
\in \{ -1,0,1 \}$.

\end{enumerate}
These $\ZZ$-gradings are the same iff the $t_i$'s differ by
signs. Furthermore, these are all good $\ZZ$-gradings of $\fs\fp_{2n}$ for
which $e(p)$ is a good element (up to conjugation by the
centralizer of $e(p)$ in $Sp_{2n}$).

\end{theorem}

\begin{corollary}
  \label{cor:5.1}
A nilpotent element $e(p)$ of $\fs\fp_{2n}$ is good for at least
one even $\ZZ$-grading iff it is either even (i.e.,~its Dynkin
grading is even), or it is odd and all even parts of $p$ have
multiplicity $2$.  It is good for at most two
even gradings.  The element $e(p)$ is good for two even gradings
if and only if all parts of $p$ are even and have multiplicity $2$; these gradings are
given by the elements $H(p\, ; \, 0,\ldots ,0)$ and $H(p\, ; \, 1,\ldots ,1)$,
the first one defining the Dynkin grading.
\end{corollary}

For each symplectic partition $p$ we construct the set $S\Pyr (p)$
of symplectic pyramids as follows. If $c(p) =0$, then $S\Pyr (p)
= \{ SP (p) \}$.  Let $c(p)>0$ and let $p_1 , \ldots , p_{c(p)}$
be all distinct even parts of $p$ which have multiplicity $2$.
If $p$ contains other parts, to each subset $\{ p_{i_1} , \ldots
,p_{i_s} \}$ we associate a symplectic pyramid obtained from $SP(p)$
by shifting by $1$ to the right (resp.~left) all boxes from the
rows corresponding to $p_{i_1} ,\ldots ,p_{i_s}$ in the (strictly)
upper (resp.~lower) half-plane.  If $p$ does not contain other
parts, there is one additional symplectic pyramid $SP_{1/2}(p)$,
obtained from $SP (p)$ by shifting all the rows in the upper
half-plane (resp.~lower half-plane) by $1/2$ to the right
(resp.~left).  We fill by basis vectors the boxes of a symplectic
pyramid and define the corresponding nilpotent element in the
same way as for $SP(p)$.  Of course, we get the same $e(p)$;
the corresponding diagonal matrix $h(SP)$ defining the good
gradation, associated with a symplectic pyramid $SP$, will have
the eigenvalue on $v_j$ equal to the first coordinate of the
corresponding box, as in the $\fgl_n$ case.

The characteristic of this good $\ZZ$-grading is obtained as
follows.  If $SP$ is different from $SP_{1/2}(p)$, we denote by
$h_i$, $0 \leq i \leq p_1 +1$, the number of boxes of $SP$ whose
first coordinates are $p_1+1-i$.  For each $h_j >0$, $0 \leq j
\leq p_1$ construct a sequence, which begins with $h_j-1$ zeros
followed by $2$ if $h_{j+1}=0$ and by $1$ if $h_{j+1}>0$.  The
number $h_{p_1+1}=2\ell$ is even (since this is the number of
boxes in the $0$\st{th} row).  Concatenating the obtained
sequences and adding $\ell$ zeros, we obtain a sequence of $n$
numbers equal to $0$, $1$ or $2$, which defines the
characteristic by assigning these integers to the corresponding
simple roots depicted by the
Dynkin diagram $\circ-\circ-\ldots-\circ-\circ
\Leftarrow \circ$.

In the case $SP=SP_{1/2}(p)$, define $h_i$ as the number of boxes
in $SP$, whose first coordinate is $p_1-i+3/2$.  We construct the
sequences for $0 \leq i \leq p_1$ as before.  The number
$h_{p_1+1}=2\ell$ is even, and we define the $p_1+1$\st{th}
sequence consisting of $2\ell -1$ zeros and one $1$.
Concatenating these $p_1+2$ sequences we obtain the
characteristic of $SP_{1/2}(p)$ (consisting of zeros and~$1$'s).

\begin{theorem}
  \label{th:5.2}
Let $\fp (a_1 ,\ldots ,a_t\, ; \, q)$ be a parabolic subalgebra
of $\fsp_{2n}$ corresponding to the pair $(a_1 , \ldots , a_t \, ;
\, q)$.  Then a Richardson element of $\fp (a_1 ,\ldots ,a_t \, ;
\, q)$ is good for the corresponding even grading of $\fsp_n$ iff
the following two properties hold:

\romanparenlist
\begin{enumerate}
\item 
$0<a_1 \leq a_2 \cdots \leq a_t$,

\item 
if $q >0$, then $a_t \leq q$  and the multiplicities of all odd $a_j$ are $\leq 1$.
\end{enumerate}
\end{theorem}

\section{Good gradings of $\fso_n$}
\label{sec:6}

Given an orthogonal partition $p$ of $N$, construct an orthogonal
pyramid $OP (p)$ as follows.  It is a centrally symmetric (around
$(0,0)$) collection of $N$ boxes of size $1 \times 1$ on the
plane, centered at points with integer coordinates (called the
coordinates of the corresponding boxes).  First, consider the
case $N=2n$ is even.  In this case the $0$\st{th} row is empty.
If $p_1$ has multiplicity $m_1$, then the rows from the $1$\st{st}
to the $[\tfrac{m_1}{2}]$\st{th} consist of boxes with the
first coordinates forming the arithmetic progression $-p_1 +1,
-p_1+3, \ldots , p_1-1$.  In the case when $m_1 =2k_1+1$ is odd, we add the $k_1+1$\st{st}
row as follows.  Let $p_i$ be the greatest part among $p_2
>p_3>\cdots$ with odd multiplicity; then the $k_1+1$\st{st} row
consists of boxes whose first coordinates form the arithmetic
progression $-p_i+1,-p_i+3,\ldots ,0,2,\ldots ,p_1-1$ (recall
that both $p_1$ and $p_i$ are odd).  After that we remove the
parts $p_1$ and $p_i$ from the partition $p$ and continue
building up the pyramid for the remaining parts.  The rows in the
lower half-plane are obtained by the central symmetry.

In the case when $N$ is odd, $OP (p)$ always contains the box
with coordinates $(0,0)$.  If the multiplicity of $p_1$ is even,
then the $0$\st{th} row contains no other boxes, and all the other
rows of the pyramid $OP (p)$ are constructed in the same way as
for $N$ even.  If the multiplicity of $p_1$ is odd, then $p_1$ is
odd, and the $0$\st{th} row consists of boxes whose first
coordinates form the arithmetic progression $-p_1+1, -p_1+3,
\ldots ,p_1-1$.  The partition $p'=(p^{m_2}_2,p^{m_3}_3, \ldots)$
is a partition of an even integer, and the non-zero rows of $OP
(p)$ simply coincide with the corresponding rows of $OP (p')$.

We fill the boxes of $OP (p)$ by basis vectors in the same way as
for $\fs\fp_N$, except that in the case of $\fs\fo_N$, $N$ odd, the vector
$v_0$ is put in the $0$\st{th} box.  The nilpotent $e(p)$,
associated to the orthogonal partition $p$, is defined as the
endomorphism which maps each basis vector to its right neighbor,
and to $0$ if there is no right neighbor,
with the following exceptions.  If $N$ is odd and the $0$\st{th}
row consists only of the box $(0,0)$ , the vector in the box with
coordinates $(-2,-\ell)$ and no right neighbor is mapped to
$v_0$, and $v_0$ is mapped to the vector in the box $(2,\ell)$
(which has no left neighbors).  If $N$ is even, then $e(p)$ is
the sum of the operator defined by horizontal maps as above and
the sum over all pairs of unequal parts $p_i$, $p_j$ with odd
multiplicities of maps which send the vector from the box of the
row corresponding to the pair $(i,j)$ with coordinates $(0,-d)$
to that with coordinates $(2,d)$ and all other basis vectors to
$0$.

\begin{lemma}
  \label{lem:6.1}
Let $p_1,\ldots ,p_{c(p)}$ be all odd parts of an orthogonal
partition $p$, having multiplicity $2$.  Define diagonal matrices
$z (t_1 ,\ldots ,t_{c(p)}) \in \fso_N$, $t_1,\ldots ,t_{c(p)} \in
\FF$, whose $i$\st{th} diagonal entry is $t_i$ if the basis
vector lies in a box of $OP (p)$ in the (strictly) upper
half-plane corresponding to the part $p_i$, and is $-t_i$ if the
basis vector lies in the centrally symmetric box, all other
entries being zero.  Then the center of the reductive part of
$\fso^{e(p)}_N$ consists of all these matrices.
\end{lemma}

Define $h(p) \in \fso_N$, using the pyramid $OP (p)$ in the same
way as for~$\fsp_N$.

\begin{theorem}
  \label{th:6.1}
In notation of Lemma~\ref{lem:6.1}, the element $H(p\, ; \, t_1
,\ldots ,t_{c(p)})\break
:=h(p)+z(t_1,\ldots ,t_{c(p)})$ defines a good
$\ZZ$-grading of $\fso_{2n+1}$ iff one of the following cases holds:

\romanparenlist
\begin{enumerate}
\item 
$1$ does not have multiplicity $2$ in $p$ and all $t_i \in \{
-1,0,1 \}$;

\item 
$1$ has multiplicity $2$ in $p$ and $p_{c(p)-1}$ is the smallest
part of $p$ not equal to $1$, $t_i \in \{ -1 ,0,1 \}$ for $1 \leq
i \leq c(p)-1$, $t_{c(p)} \in \ZZ$, and
$|t_{c(p)-1}-t_{c(p)}| \leq p_{c(p)-1}-1$,
$|t_{c(p)-1}+t_{c(p)}|\leq p_{c(p)-1}-1$;

\item
 $1$ has multiplicity $2$ in $p$ and $p_{c(p)-1}\neq q$, where $q$ is the
  smallest part of $p$ greater than $1$, $t_i \in \{ -1,0,1 \}$
  for $1 \leq i \leq c(p)-1$, $t_{c(p)} \in \ZZ$ and
  $|t_{c(p)}| \leq q-1$.

\end{enumerate}

These gradings are the same iff the $t_i$ differ by signs.
Furthermore, these are all good $\ZZ$-gradings of $\fs\fo_{2n+1}$ for
which $e(p)$ is a good element (up to conjugation by the
centralizer of $e(p)$ in $SO_{2n+1}$).
\end{theorem}

\begin{theorem}
  \label{th:6.2}
In notation of Lemma~\ref{lem:6.1}, let $p$ be an ortogonal partition of
 $2n$ and let  $C(p)=\{  p_1> \ldots > p_{c(p)}\}$ be all distinct odd parts
of $p$ with multiplicity $2$.  The element $H(p\, ; \, t_1
,\ldots ,t_{c(p)}):=h(p)+z(t_1,\ldots ,t_{c(p)})$ defines a good
$\ZZ$-grading of $\fs\fo_{2n}$ with nilpotent $e(p)$ iff one of the following cases holds:
\romanparenlist
\begin{enumerate}

\item 
 $1\notin C(p)$, there is a part $p_i$ of $p$ such that $p_i\notin
 C(p)$ and all $t_i \in \{-1,0,1 \}$;
\vskip 0pt plus 4pt minus 2pt

\item 
$1\in C(p)$, there is a part $p_i$ of $p$ such that $p_i\notin C(p)$
and $p_{c(p)-1}$ is the smallest part of $p$ not equal to $1$,
$t_i \in \{ -1 ,0,1 \}$ for $1 \leq
i \leq c(p)-1$, $t_{c(p)} \in \ZZ$, and
$|t_{c(p)-1}-t_{c(p)}|\leq p_{c(p)-1}-1$,
$|t_{c(p)-1}+t_{c(p)}|\leq p_{c(p)-1}-1$;
\vskip 0pt plus 4pt minus 2pt

\item
$1\in C(p)$, there is a part $p_i$ of $p$ such that $p_i\notin C(p)$
and $p_{c(p)-1} \neq  q$, the smallest part of $p$ greater than $1$,
$t_i \in \{ -1 ,0,1 \}$ for $1 \leq i \leq c(p)-1$, $t_{c(p)} \in \ZZ$,
and $|t_{c(p)}| \leq q-1$;
\vskip 0pt plus 4pt minus 2pt

\item 
 $1\notin C(p)$, all parts $p_i$ of $p$ lie in $C(p)$, $2t_i,
 t_i+t_j, t_i-t_j \in \ZZ$, $ -2 \leq 2t_i\leq 2$
 for $1 \leq i< j \leq c(p)$;
\vskip 0pt plus 4pt minus 2pt

\item 
$1\in C(p)$, all parts $p_i$ of $p$ lie in $C(p)$,
$2t_i, t_i+t_j, t_i-t_j \in \ZZ$, for $1 \leq i< j \leq c(p)$
$ -2 \leq 2t_i\leq 2$ for $1 \leq i< j \leq c(p)-1$,
$| t_{c(p)-1}-t_{c(p)}|\leq p_{c(p)-1}-1$,
$| t_{c(p)-1}+t_{c(p)}|\leq p_{c(p)-1}-1$.

\end{enumerate}

These gradings are the same iff the $t_i$ differ by signs.
Furthermore, these are all good $\ZZ$-gradings of $\fs\fo_{2n}$ for
which $e(p)$ is a good element (up to conjugation by the
centralizer of $e(p)$ in $O_{2n}$).
\end{theorem}

Next, for each orthogonal partition $p$ of $N$ we construct the
set $OPyr (p)$ of orthogonal pyramids.  If $c(p)=0$, then
$OPyr(p)= \{ OP (p) \}$.  Assume now that $c(p) >0$,
and let $C(p)=\{  p_1> \ldots > p_{c(p)}\}$ be all distinct odd parts of $p$ with
multiplicity $2$.

First consider the case of $N$ odd. If $1\notin C(p)$, to each
subset $\{ p_{i_1}, \ldots ,p_{i_s}\}$ of $\{ p_1 ,\ldots
,p_{c(p)}\}$ we associate an orthogonal pyramid obtained from $OP
(p)$ by shifting by $1$ to the right (resp.~left) all boxes from
the rows corresponding to $p_{i_1}, \ldots ,p_{i_s}$, in the
strictly upper (resp.~lower) half-plane. If $1\in C(p)$,
i.e.,~$p_{c(p)}=1$, let $q$ be the smallest part of the
partition $p$, which is greater than $1$.  If $q$ is not an odd
part with multiplicity $2$, then for each pair $(\{ p_{i_1},\ldots
,p_{i_s}\} \, , \, t_{c(p)})$, where $\{ p_{i_1} , \ldots , p_{i_s} \}
\subset \{ p_1,\ldots ,p_{c(p)-1}\}$ and $0 \leq t_{c(p)}\leq
q-1$, $t_{c(p)} \in \ZZ$, we construct an orthogonal pyramid
obtained from $OP (p)$ by shifting by $1$ to the right
(resp.~left) all boxes from the rows corresponding to
the parts $p_{i_1},\ldots,p_{i_s}$, and the box corresponding to part $1$, by
$t_{c(p)}$ to the right (resp.~left) in the strictly upper
(resp.~lower) half plane.  Under these assumptions we obtain
$2^{c(p)-1} q$ orthogonal pyramids, which form the set $OPyr
(p)$.  If, finally, $q=p_{c(p)-1}$ and $p_{c(p)} =1$, then for
each pair $(\{ p_{i_1}, \ldots ,p_{i_s}\} \, , \, t_{c(p)})$
(resp.~$(\{ p_{i_1} ,\ldots ,p_{i_s}\, , \,  p_{c(p)-1}\}\, , \,
t_{c(p)})$), where $\{p_{i_1},\ldots ,p_{i_s} \} \subset \{ p_1,\ldots
,p_{c(p)-2}\}$ and $0 \leq t_{c(p)} \leq p_{c(p)-1}-1$ (resp.~$0
\leq t_{c(p)}\leq p_{c(p)-1}-2$), $t_{c(p)}\in \ZZ$, we construct
an orthogonal pyramid as in the previous case.  We thus obtain
$2^{c(p)-2} (2q-1)$ orthogonal pyramids, which form the set
$OPyr (p)$.

Let now $N$ be even.  If $p$ has parts which are not in $C(p)$, then
$OPyr (p)$ is constructed in the same way as for $N$ odd.
Finally, suppose all parts of $p$ are odd of multiplicity $2$.
If $1 \notin C(p)$, then $O\Pyr (p)$ is constructed in the same way
as $SPyr (p)$.  If $1 \in C(p)$, then $OPyr (p)$ consists of the
pyramids constructed as for $N$ odd and the set of pyramids
$OP_{1/2, p_{c(p)-1}}(p)$ obtained by a shift by
$1/2$ to the right (resp.~left) of all rows
corresponding to the parts $p_1 ,\ldots,p_{c(p)-1}$ and the box
from the row, corresponding to $1$, by $t$ to the right
(resp.~left) in the upper (resp.~lower) half-plane, where $t \in
\tfrac{1}{2}+\ZZ$, $\tfrac{1}{2}\leq t \leq p_{c(p)-1}-\tfrac{3}{2}$.

The characteristic of this good $\ZZ$-grading is obtained as
follows.  If $OP$ is different from $OP_{1/2, p_{c(p)-1}}(p)$, we denote by
$h_i$, $0 \leq i \leq p_1 +1$, the number of boxes of $SP$ whose
first coordinates are $p_1+1-i$.  For each $h_j >0$, $0 \leq j
\leq p_1$ construct a sequence, which begins with $h_j-1$ zeros
followed by $2$ if $h_{j+1}=0$ and by $1$ if $h_{j+1}>0$.  The last
number $h_{p_1+1}=2\ell$ is even for even $N$ and  $h_{p_1+1}=2\ell+1$ is odd
for odd $N$ (since this is the number of boxes in the $0$\st{th} row). If for
even $N$, $h_{p_1+1}=0$, then $h_{p_1}=2\ell$ is even and
positive and we define the  $p_1+1$\st{th} sequence  consisting of $2\ell -1$
zeros and one $2$.
Concatenating the obtained sequences and adding $\ell$ zeros, we obtain
a sequence of $n$
numbers equal to $0$, $1$ or $2$, which defines the
characteristic by assigning these integers to the corresponding
simple roots depicted by the
Dynkin diagram $\begin{smallmatrix} &&&&&&
    \circ\\
&&&&&&|\\
\circ&-&\circ&-&\cdots&-&\circ&-&\circ
\end{smallmatrix}$ for even $N$ and Dynkin diagram
$\circ-\circ-\cdots-\circ-\circ
\Rightarrow \circ$ for odd $N$   .

In the case  $OP_{1/2, p_{c(p)-1}}(p)$, define $h_i$ as the number of boxes
in $SP$, whose first coordinate is $p_1-i+3/2$.  We construct the
sequences for $0 \leq i \leq p_1$ as before.  The number
$h_{p_1+1}=2\ell$ is even, and we define the $p_1+1$\st{th}
sequence consisting of $2\ell -1$ zeros and one $1$.
Concatenating these $p_1+2$ sequences we obtain the
characteristic of  $OP_{1/2, p_{c(p)-1}}(p)$ (consisting of zeros and $1$'s).

The ortogonal partition of an odd number $N$ must have an odd number
of odd
parts with odd multiplicity. Hence, if the nilpotent $e(p)$ is even
then all parts of $p$ are odd and the Dynkin grading defined by $h(p)$ is an
even grading for which $e(p)$ is good. If
$e(p)$ is good for a non-Dynkin grading then, by Theorem~\ref{th:1.1},
we have $c(p)>0$. Therefore,
by Theorem~\ref{th:6.1},
the semisimple element $H(p\,;t_1,\ldots,t_{c(p)})$ is even
if and only if the partition $p$ is even. From this we obtain
\begin{lemma}
  \label{lem:6.2}
The nilpotent $e(p)$ of $\fso_{N}$, $N$ odd,
for a partition $p$ of $N$ is good for an
even grading iff all parts $p_i$ are odd.
\end {lemma}

By Lemma~\ref{lem:6.2} and Theorem~\ref{th:6.1} the nilpotent $e(p)$
can be good for even non-Dynkin gradings
if and only if $1\in C(p)$ and all parts $p_i$
of $p$ are odd. We now apply the above algorithm
for constructing the characteristic of the grading
for an orthogonal pyramid from $OPyr(p)$ to obtain
\begin{theorem}
  \label{th:6.3}
  Let $p = \fp (a_1 ,\ldots ,a_t\, ; \, q)$ be a parabolic subalgebra
  of $\fso_{2n+1}$ corresponding to the pair $(a_1 , \ldots , a_t
  \, ; \, q)$ (recall that in this case $q$ is odd).  Then a
  Richardson element of $\fp (a_1 ,\ldots ,a_t
  \, ;\, q)$ is good for the corresponding even grading of
  $\fso_{2n+1}$ iff one of the following properties holds:

\romanparenlist
\begin{enumerate}
\item 
$0<a_1 \leq a_2 \cdots \leq a_t \leq q$;
\vskip 0pt plus 4pt minus 2pt

\item 
$p$ has the form $\fp (a_1 , \ldots ,  a_{t-2},
a_{t-1}^{m_{t-1}-s},  a_t,  a_{t-1}^s ;  q)$ , where \newline
$0 < a_1 \leq \cdots \leq a_{t-2} <  a_{t-1}= q< a_t= q+1$ and $0\leq s \leq m_{t-1}$;
\vskip 0pt plus 4pt minus 2pt

\item 
$0 < a_1 \leq a_2 \cdots \leq a_{t-1}  < q <a_t=q+1 $ .
\end{enumerate}
\end{theorem}

\begin{lemma}
  \label{lem:6.3}
The nilpotent $e(p)$ of $\fs\fo_{2n}$ for a partition $p$ of $2n$
is good for an
even grading iff one of the two following properies holds:
\romanparenlist
\begin{enumerate}
\item 
 $p$ is an even partition;
\vskip 0pt plus 4pt minus 2pt

\item 
if $ p$ is not even then all odd parts $p_i$ of $p$ have
multiplicity $m_i=2$.

\end{enumerate}

\end{lemma}

By Lemma~\ref{lem:6.3} and Theorem~\ref{th:6.2} the nilpotent $e(p)$
can be good for an even non-Dynkin grading
iff $1\in C(p)$ and all parts $p_i$
of $p$ are odd. We now apply the above algorithm
for constructing the characteristic of the grading
for an orthogonal pyramid from $OPyr(p)$ to obtain

\begin{theorem}
  \label{th:6.4}
Let $\fp =\fp (a_1 ,\ldots ,a_t; \, q)$ be a parabolic
subalgebra of $\fso_{2n}$ corresponding to the pair $(a_1,
\ldots , a_t; \, q)$ (recall that in this case $q$ is
even $\neq 2$). Then a Richardson element of $\fp (a_1,\ldots
,a_t;\, q)$ is good for the corresponding even grading of
$\fso_{2n}$ iff one of the following  properties holds:

\romanparenlist
\begin{enumerate}
\item 
$n=2s+1,q=0$ and composition $(a_1,\ldots ,a_t)$ has
one of the following forms:
\newline $(1^{2(s-i)}, 2^i, 1)$, where $i=0,1, \ldots ,s$;
\newline $(1^{2(s-i)+1}, 2^{i-l}, 3, 2^{l-2}, 1)$, where $i=2, \ldots , s, l=2,\ldots ,i$;
\newline $(2^{s-(i+1)}, 3,  2^i)$, where $i=1,\ldots, s-1$;
\vskip 0pt plus 4pt minus 2pt

\item 
$n=2s+2,q=0$ and composition $(a_1^{m_1} ,\ldots ,a_t^{m_t})$ has
one of the fol\-lowing forms:
\newline $(1^{2(s-i)+1}, 2^i, 1)$, where $i=0,1,\ldots ,s$;
\newline $(1^{2(s-i)}, 2^{i-l}, 3, 2^{l-1}, 1)$, where $i=1, \ldots , s, l=1,\ldots ,i$;
\newline $(1,2^{(i-1)}, 3, 2^{s-i})$, where $i=1,\ldots, s-1$;

\item 
$q=0$, $0<a_1 \leq a_2 \cdots\leq a_t $, and multiplicities of odd $a_j$ are at most 1;

\item 
 $q=0$, $0<a_1 \leq a_2 \cdots \leq a_{t-1}$, $a_t = a_{t-1}
-1>0$, where $a_{t-1}$ is odd, $a_{t-1} \geq 5$, $m_t >0$, and multiplicities
of odd $a_j$ are at most 1; \vskip 0pt plus 4pt minus 2pt

\item 
$q>0$, $0<a_1 \leq a_2 \cdots \leq a_t \leq q$; \vskip 0pt plus 4pt minus
2pt

\item 
$\fp$ has the form $\fp \, (a_1 \, ,\, \ldots \, , \, a_{t-2} \, ,
\, a_{t-1}^{m_{t-1}-s} \, , \, a_t \, , \, a_{t-1}^s \, ; \; q)$,
where\break $0<a_1 \leq a_2 \cdots \leq a_{t-2} <  a_{t-1}= q< a_t= q+1$, $
m_{t-1}>0$, and $0\leq s
\leq m_{t-1}$; \vskip 0pt plus 4pt minus 2pt

\item 
$0\leq a_1 \leq a_2 \cdots \leq a_{t-1}  < q <a_t=q+1 $ .
\end{enumerate}
\end{theorem}

\section{Good gradings of exceptional simple Lie algebras}
\label{sec:7}

If $\fg$ is an exceptional simple  Lie algebra $G_2$ or $F_4$,
then, due to the tables of \cite{E}, the reductive part of
$\fg^e$ is semisimple (or zero) for any nilpotent element $e$ of
$\fg$.  Hence, by Corollary~\ref{cor:1.1}, we have

\begin{theorem}
  \label{th:7.1}
All good $\ZZ$-gradings of $G_2$ and $F_4$ are the Dynkin ones.
\end{theorem}

Due to the tables of \cite{E} for the exceptional simple Lie
algebras $E_6$, $E_7$ and $E_8$ the number of nilpotent orbits
for which the reductive part of the centralizer is not semisimple
(and not zero) is $10$, $6$ and $7$, respectively (out of $21$,
$45$ and $70$, respectively).  In notation of \cite{C} they are
as follows:
\begin{eqnarray*}
  E_6 \!&\!:\!&\!
      D_5, D_5(a_1), A_4 \!+\! A_1, D_4 (a_1), A_4, A_3 \!+\! A_1, A_3,
      A_2 \!+\! 2A_1, A_2 \!+\! A_1, 2A_1 \, ;\\
  E_7 \!&\!:\!&\! E_6 (a_1), D_5 (a_1), A_4+A_1, A_4, A_3+A_2, A_2+A_1 \,;\\
  E_8 \!&\!:\!&\! D_7 (a_1), E_6 (a_1)+A_1, D_7 (a_2), D_5 +A_2,
          A_4+2A_1,A_4+A_1, A_3+A_2 \, .
\end{eqnarray*}

We use Theorem~\ref{th:1.1} in order to describe all good
$\ZZ$-gradings for which the above listed nilpotent elements are
good.  The answer is given in Tables~$E_6$, $E_7$ and $E_8$,
where we list all nilpotent orbits which admit non-Dynkin good
gradings.  (The Dynkin gradings are described in \cite{D},
\cite{E},~\cite{C}.)

\begin{table}[tbp]
  \centering
  \vskip 2pt plus 8pt minus 5pt
  \caption{$E_6$}
\begin{tabular}{cc|cc}
\hline\\[-2ex]
 $D_5$ &
$
\begin{array}{ccccc}
2&0&2&0&2\\&&2
\end{array}
$ & $
\begin{array}{ccccc}
2&1&1&1&1\\&&2
\end{array}
$
&
$
\begin{array}{ccccc}
2&2&0&2&0\\&&2
\end{array}
$\\
&&
$
\begin{array}{ccccc}
2&2&1&1&1\\&&1
\end{array}
$
&
$
\begin{array}{ccccc}
2&2&2&0&2\\&&0
\end{array}
$\\
\hline\\[-2ex]
 $D_5 (a_1)$
&
$
\begin{array}{ccccc}
1&1&0&1&1 \\ &&2
\end{array}
$
&
$
\begin{array}{ccccc}
2&0&0&2&0 \\ && 2
\end{array}
$\\
\hline\\[-2ex]
$A_4 + A_1$ &
$
\begin{array}{ccccc}
1&1&0&1&1 \\ &&1
\end{array}
$&

$
\begin{array}{ccccc}
2&0&2&0&0 \\ &&0
\end{array}
$\\
\hline\\[-2ex]
$D_4 (a_1)$ &
$
\begin{array}{ccccc}
0&0&2&0&0 \\&&0
\end{array}
$ &
$
\begin{array}{ccccc}
1&0&1&1&0 \\&&0
\end{array}
$&
$
\begin{array}{ccccc}
1&1&0&1&1 \\ &&0
\end{array}
$
\\
&&
$
\begin{array}{ccccc}
2&0&0&2&0 \\&&0
\end{array}
$\\
\hline\\[-2ex]
 $A_4$ &
$
\begin{array}{ccccc}
2&0&0&0&2 \\&&2
\end{array}
$&
$
\begin{array}{ccccc}
2&1&0&1&0 \\ &&1
\end{array}
$&
$
\begin{array}{ccccc}
2&0&0&2&2 \\ &&0
\end{array}
$\\
\hline\\[-2ex]
 $A_3 + A_1$ &
$
\begin{array}{ccccc}
0&1&0&1&0 \\ &&1
\end{array}
$&
$
\begin{array}{ccccc}
1&1&0&0&1 \\&&1
\end{array}
$&
$
\begin{array}{ccccc}
2&0&1&0&1 \\&&0
\end{array}
$\\
\hline\\[-2ex]
 $A_3$ &
$
\begin{array}{ccccc}
1&0&0&0&1\\ &&2
\end{array}
$&
$\begin{array}{ccccc}
2&0&0&0&0 \\&&2
\end{array}
$&
$
\begin{array}{ccccc}
2&1&0&0&0 \\ && 1
\end{array}
$\\
&&
$
\begin{array}{ccccc}
2&2&0&0&0 \\&&0
\end{array}
$\\
\hline\\[-2ex]
 $A_2 + 2A_1$ &
$
\begin{array}{ccccc}
0&1&0&1&0 \\&& 0
\end{array}
$&
$
\begin{array}{ccccc}
0&2&0&0&0 \\&&0
\end{array}
$\\
\hline\\[-2ex]
 $2A_1$
&
$
\begin{array}{ccccc}
1&0&0&0&1 \\&& 0
\end{array}
$&
$
\begin{array}{ccccc}
2&0&0&0&0 \\&& 0\\
\end{array}
$\\
\hline
  \end{tabular}
  \vskip 3pt plus 8pt minus 5pt
\end{table}

\begin{theorem}
  \label{th:7.2}
All good non-Dynkin $\ZZ$-gradings of $E_6$, $E_7$ and $E_8$ are
described in the second column of Tables~$E_6$, $E_7$ and $E_8$,
respectively, where we list all good non-Dynkin $\ZZ$-gradings
for which the nilpotent of the first column is good.  In
Table~$E_6$ the characteristics of good non-Dynkin $\ZZ$-gradings
are listed up to the symmetry of the Dynkin diagram.
\end{theorem}

\begin{table}[tbp]
  \vskip 3pt plus 8pt minus 5pt
  \centering
  \caption{$E_7$}
  \begin{tabular}{cc|ccc}\hline\\[-2ex]
 $E_6 (a_1)$ &
$\begin{array}{cccccc}
0&2&0&2&0&2\\&&&0
\end{array}
$
&
$
\begin{array}{cccccc}
1&1&1&1&0&2 \\&&&1
\end{array}
$
&
$
\begin{array}{cccccc}
2&0&2&0&0&2 \\&&&2
\end{array}
$\\
&&
$
\begin{array}{cccccc}
2&1&1&0&1&1 \\&&&2
\end{array}
$ &
$\begin{array}{cccccc}
2&2&0&0&2&0 \\ &&&2
\end{array}
$\\
\hline\\[-2ex]
 $D_5 (a_1)$&
$
\begin{array}{cccccc}
0&1&0&1&0&2 \\&&& 0
\end{array}
$
&$
\begin{array}{cccccc}
2&0&0&0&0&2 \\&&&2
\end{array}
$\\
\hline\\[-2ex]
 $ A_4 + A_1$ &
$
\begin{array}{cccccc}
0&1&0&1&0&1 \\&&& 0
\end{array}
$ &$
\begin{array}{cccccc}
1&0&1&0&0&1 \\&&&1
\end{array}$
&
$
\begin{array}{cccccc}
2&0&2&0&0&0 \\ &&& 0
\end{array}
$\\ &&
$
\begin{array}{cccccc}
2&0&1&0&1&0 \\&&& 0
\end{array}
$&
$\begin{array}{cccccc}
2&0&0&0&2&0 \\&&& 0
\end{array}$\\
\hline\\[-2ex]
 $A_4$ &
$
\begin{array}{cccccc}
0&2&0&0&0&2 \\ &&& 0
\end{array}
$&$
\begin{array}{cccccc}
2&1&0&0&0&1 \\ &&& 1
\end{array}
$\\
\hline
 \end{tabular}
  \vskip 3pt plus 8pt minus 5pt

\end{table}

Note that among the $10$ nilpotent orbits of $E_6$ there is
exactly one for which the reductive part is not semisimple, but
the Dynkin grading is the only good one (and for $E_7$ and $E_8$
there are $2$ and $5$ such nilpotents, respectively).  This is
the nilpotent of type $A_2+A_1$.  Pick one such nilpotent $e$,
and let $h(e)$ be the semisimple element which defines the
corresponding Dynkin grading of $E_6$.  The reductive part of its
stabilizer is a direct sum of $A_2$ and a $1$-dimensional torus
$T_1$ \cite{E}.  Among the weights of $h(e)+T_1$ on $E^e_6$ there
are weights $1 \pm t$ and $1+3t$.  Hence $t \in \ZZ$ and the
condition that $(E^e_6)_j =0$ for $j<0$ gives the inequalities
$-\tfrac{1}{3} \leq t \leq \tfrac{1}{3}$.  Hence $t=0$.  Similar
argument is used to describe all good $\ZZ$-gradings for all
other nilpotents as good elements.

Let us give one more example---the nilpotent $e$ of type $A_4$ in
$\fg =E_6$.  It is easy to see that for the corresponding Dynkin
$\ZZ$-grading of $E_6$ one has:
\begin{displaymath}
  \dim \fg_0 =18 \, , \,\, \dim \fg_2 =14 \, , \,\,
  \dim  \fg_4=9\, , \,\, \dim \fg_6 =6 \, , \,\, \dim \fg_8=1 \, .
\end{displaymath}
Hence, by (0.2), we have:
\begin{displaymath}
  \dim \fg^e_0 =4 \, , \,\, \dim \fg^e_2 =5 \, , \,\,
   \dim \fg^e_4=3 \, , \,\, \dim \fg^e_6=5 \, , \,\,
    \dim \fg^e_8 =1 \, .
\end{displaymath}
The reductive part of $\fg^e$ is $\fg^e_0 \simeq A_1 \oplus T_1$,
where $T_1 =\{ t|t \in \FF \}$ is the $1$-dimensional torus.
{}From the explicit description of the basis of $\fg^e$, which can
be obtained by using the explicit form of $e$ in \cite{C}, \cite{D}, \cite{E}, and
the computer program GAP  (see
www-gap.dcs.st-and.ac.uk/$\sim$gap/), we find the
weights of $h(e) +T_1$ on $\fg^e_j$ $(j \in 2\ZZ_+)$:  $2 \pm 3t$
and $2;4 \pm 6t$ and $4; 6 \pm 3t;6;8$.  The conditions of
integrability and non-negativity of a good $\ZZ$-grading give the
following restrictions:
\begin{displaymath}
  3t \in \ZZ \, , \quad -2 \leq 3t \leq 2 \, .
\end{displaymath}
Hence $t=0$, $\tfrac{1}{3}$, $\tfrac{2}{3}$, $-\tfrac{1}{3}$, $-\tfrac{2}{3}$
give all good $\ZZ$-gradings with $e$ a good element.  Of course,
$t=0$ gives the Dynkin grading, and it is easy to see that
$t=\tfrac{1}{3}$ and $\tfrac{2}{3}$ give the good gradings in the
line $A_4$ of Table~$E_6$.  The values $t=-\tfrac{1}{3}$ and
$-\tfrac{2}{3}$ give the good $\ZZ$-gradings obtained by the
symmetry of the Dynkin diagram.

\begin{table}[btp]
  \vskip 4pt plus 8pt minus 5pt
  \centering
   \caption{$E_8$}
  \begin{tabular}{cc|ccc}\hline\\[-2ex]
 $D_7 (a_1)$ &$
\begin{array}{ccccccc}
2&0&0&2&0&0&2 \\ &&&& 0
\end{array}
$&
$
\begin{array}{ccccccc}
1&1&1&0&0&1&1 \\ &&&&1
\end{array}
$\\
\hline\\[-2ex]
 $D_7 (a_2)$ & $
\begin{array}{ccccccc}
1&0&1&0&1&0&1 \\ &&&& 0
\end{array}
$ &
$
\begin{array}{ccccccc}
0&2&0&0&0&2&0 \\ &&&& 0
\end{array}
$\\
\hline
 \end{tabular}
  \vskip 3pt plus 8pt minus 5pt

\end{table}


\end{document}